\documentstyle[11pt,paspconf,epsf]{article}

\begin{document}
\title{Comparing Low Surface Brightness Galaxies and Ly$\alpha$
Absorbers}
\author{Suzanne M. Linder}
\affil{Department of Astronomy and Astrophysics, The Pennsylvania 
State University, 525 Davey Laboratory, University
Park, Pennsylvania, 16802 USA}

\begin{abstract}
I explore the hypothesis that Ly$\alpha$ absorption at low redshift is
caused by the outer regions of extended galaxy disks, including those of
dwarf and low surface brightness galaxies.  McGaugh (1996) has shown that
the distribution of central galaxy surface brightnesses may be flat to
25 B mag arcsec$^{-2}$ or fainter.  A population of galaxies is simulated
based upon observed distributions.  The low-redshift neutral column 
density distribution is predicted, and an estimate is made for the number
density of disk galaxies required to explain all absorbers. 
Results relating to galaxy luminosities and impact parameters
are discussed.
\end{abstract}


Determining the relation between galaxies and 
Ly$\alpha$  forest absorbers is complicated by uncertainties in the extent and
configuration of galactic gas as well as uncertainties in the properties 
of the galaxy
population.  The possibility is explored that forest absorbers 
at low redshift can be explained by highly extended, outer galaxy disks,
where galaxies obey a flat central surface brightness 
distribution.  This model is likely to represent field galaxies best, as
interactions and clustering are not considered.  Conclusions from various
studies depend upon the column density threshold, where generally lower 
column density absorbers are less likely to be associated with galaxies
(Lanzetta et al. 1995; le Brun, Bergeron, \&Boiss\'e 1996;
Morris et al. 1993; Bowen, Blades, and Pettini 1996).

Galaxy disks are observed to have sharp cutoffs at neutral column densities 
of a 
few times $10^{19}$  cm$^{-2}$, which have been explained by rapidly
increasing ionization (Maloney 1993; Corbelli \& Salpeter 1993; Dove \&
Shull 1994).  Ionized disks may extend to large radii
and allow substantial cross sections for low column density
absorption.  While distant gas associated with galaxies
may not behave as perfect, rotating disks, cosmological simulations
indicate that galaxies form within extended sheets.
Dinshaw et al. (1995) used double lines of
sight to show that absorbers are hundreds of kpc in size.  Furthermore,
the observations were found to be consistent with flattened absorbers 
having organized motion, such as rotating disks (Charlton, Churchill, \& Linder
1995).

A population of low surface
brightness (LSB) galaxies has been discovered recently.  For example,
the Schombert et al. (1992) catalog contains $\sim$340.
About ninety percent of these resemble high surface brightness galaxies
in size and other properties.  The remaining LSB galaxies are highly 
extended Malin
objects, with disk scale lengths of $\sim$50 kpc or more.
While Malin objects
are extremely low in surface brightness ($\sim$26 mag arcsec$^{-2}$), they
are often among the most luminous galaxies known.  
The Freeman law ($\mu_0=21.65\pm 0.35$ B mag arcsec$^{-2}$)
was shown to result from selection effects.  McGaugh 
(1996) found that the surface brightness distribution is approximately
flat to at least 24 mag arcsec$^{-2}$, with only a rapid 
falloff brightward.
Most LSB galaxies discovered
to date are rich in gas, so they must make a contribution to
Ly$\alpha$ absorption.  Since the number density of extremely low surface
brightness galaxies is highly uncertain, it is useful to constrain it
using Ly$\alpha$ absorber counts.

A population of disk galaxies is defined using observed 
distributions.  For a simulated sample,
each disk is modeled using pressure and gravity 
confinement (Charlton, Salpeter \& Hogan 1993).
The galaxies are given random locations,
and lines of sight are chosen.  Neutral
column densities are found by integrating densities along
lines of sight through galaxies.  
A flat surface brightness distribution is assumed following McGaugh (1996).
The cutoff at the LSB end is highly uncertain, although objects have
been observed with $\mu_0>27$ mag arcsec$^{-2}$.  A scale length-surface
brightness relation is assumed based upon Sprayberry et al. (1995).
The model utilizes a scale length distribution from de Jong \& van der 
Kruit (1994).  Rotation velocities are found using the Tully-Fisher relation
(Zwaan et al. 1995). Surface brightness is assumed to be related to 
central column density
(de Blok and McGaugh 1996).
An extrapolated central column density for exponential disks (Bowen, Blades
\& Pettini 1995)
is assumed to correspond to the Freeman surface brightness, and a 
surface brightness-column density relation
is derived based upon McGaugh (1996).

The total (neutral plus ionized) column density is assumed to behave
as $N_{tot}(r)=N_0\exp(-r/h_r)$ where $r<r_{sw}$ and $N_0\exp(r_{sw}/
h_r)(r_o/r_{sw})^{-p}$ for $r>r_{sw}$.  The ratio of $h_r$ to the
B scale length is uncertain, but an initial estimate of 3.4 (Salpeter 1995)
is derived.  Power law behavior in outer disks is required by 
that observed in the high redshift column
density distribution, which is found to
behave as $N_H^{-1.5}$ (Hu et al. 1995), corresponding to $p=4$.
The location of the power law switch is assumed to occur at the
same location as the ionization transition, at 
$3\times 10^{19}$ cm$^{-2}$.  These two transitions are likely to be
unrelated, however.  The latter occurs at 
fairly constant $N_H$, but the location of the power law 
switch is uncertain.  Evidence is observed for such a 
transition in dwarfs by Hoffman et al. (1993).  The ionization level 
is given by Kulkarni and Fall (1993).  
Using the pressure and gravity confinement model requires assuming
an external pressure, which may be due to infalling
material or a local or diffuse intergalactic medium
(Charlton, Salpeter, \& Linder
1994).

\begin{figure}
\plottwo{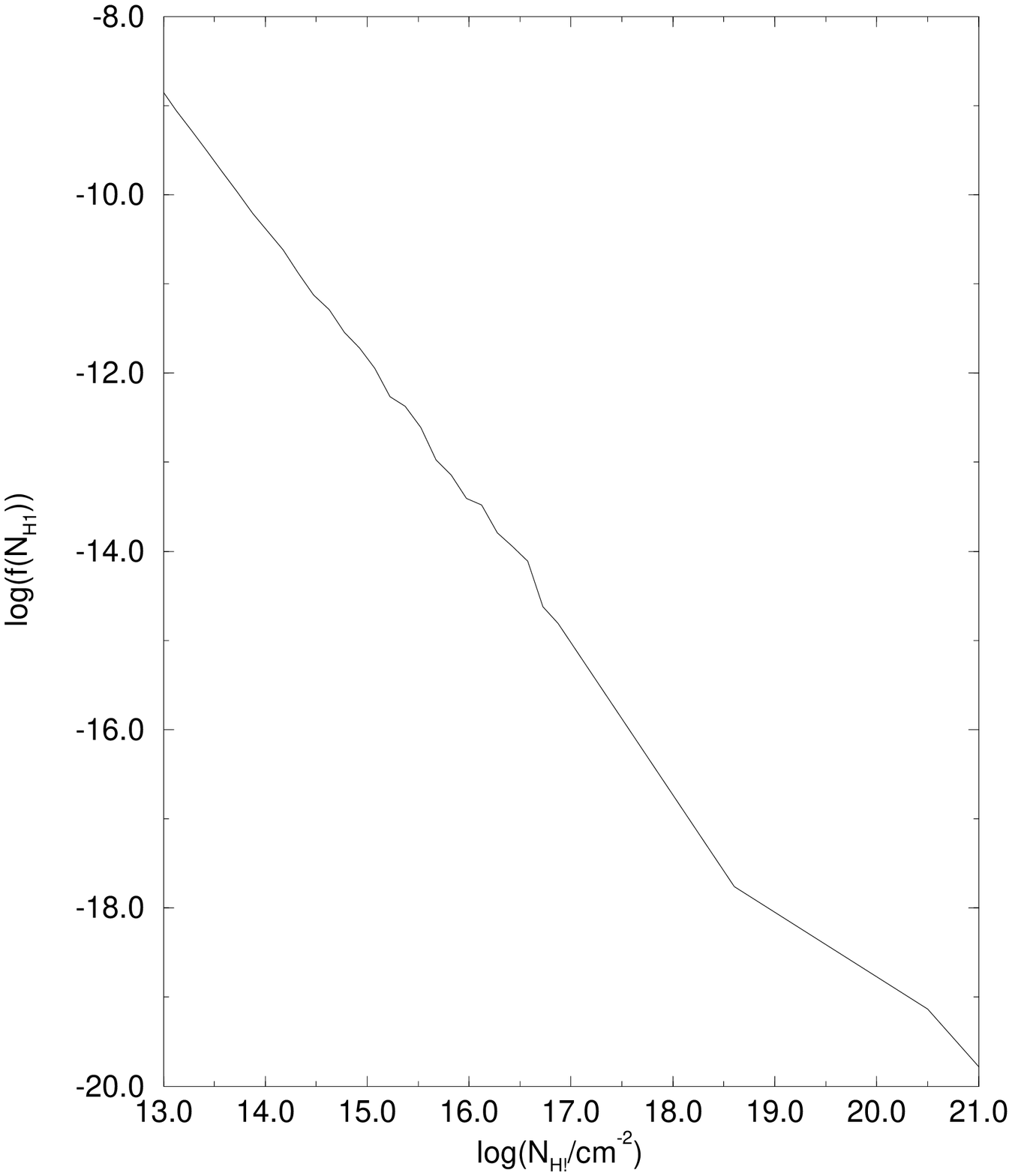}{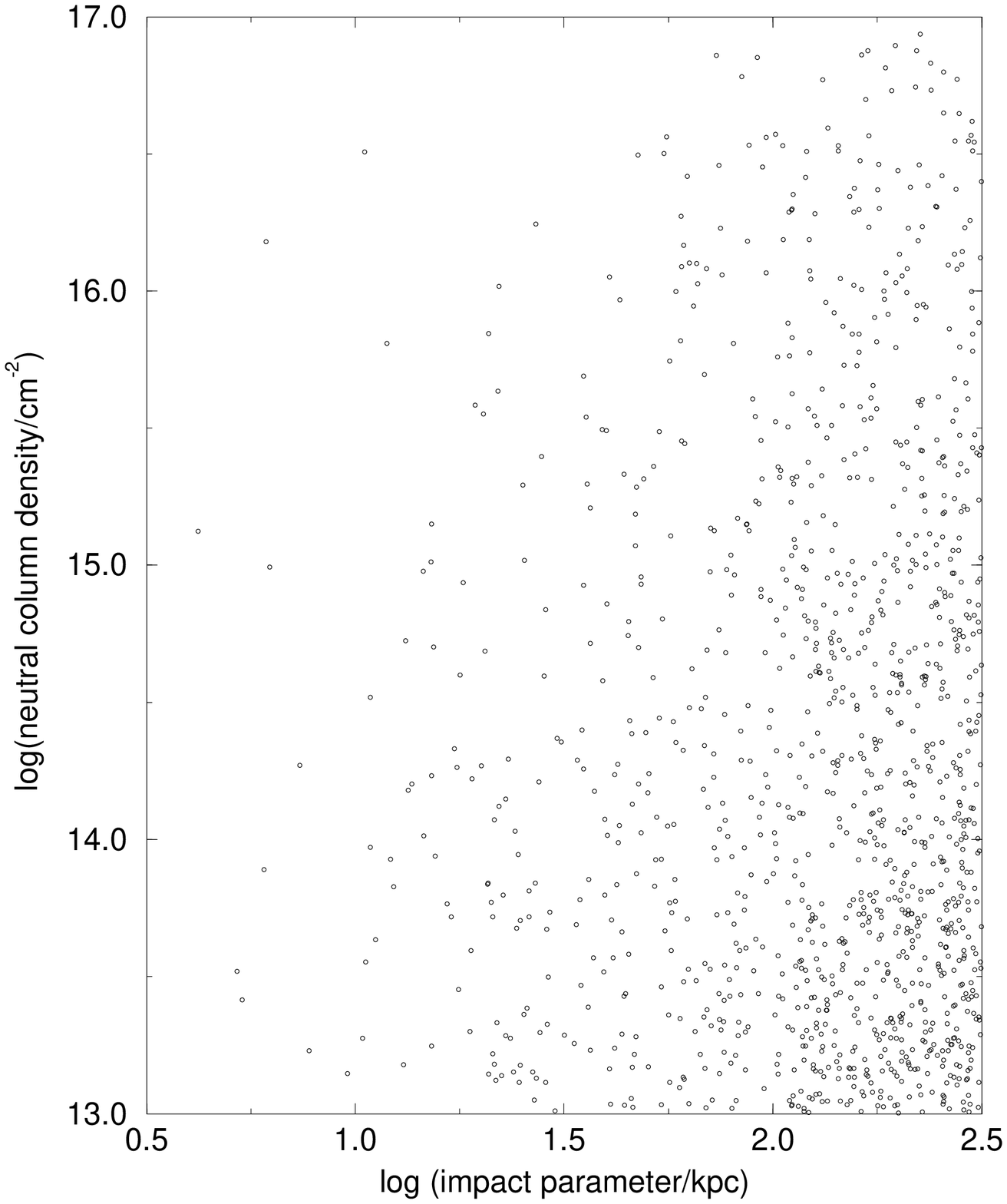}
\caption{a) The predicted low redshift column density distribution.
b) Column density versus galaxy impact parameter.  For a single galaxy,
column density would fall off with a slope of -4.}\label{fig-1}
\end{figure}

The model low redshift column density distribution is
in Figure~\ref{fig-1}a.
Model parameters
will be constrained when an accurate observed distribution is 
available.  The 
number density of disk galaxies required to reproduce observed absorber
counts (Bahcall et al. 1996) is estimated as 0.3 Mpc$^{-3}$.  Neutral
column density values are plotted versus impact parameter in
Figure~\ref{fig-1}b.
An anticorrelation is seen only when looking out several Mpc
due to large cross sections of luminous galaxies.
Any anticorrelation is weakened by disk inclination effects and by
varying absorption radii.  Absorption radii in this model
are strongly dependent
upon galaxy luminosity, but LSB luminosities are routinely underestimated
observationally.

\begin{figure}
\plottwo{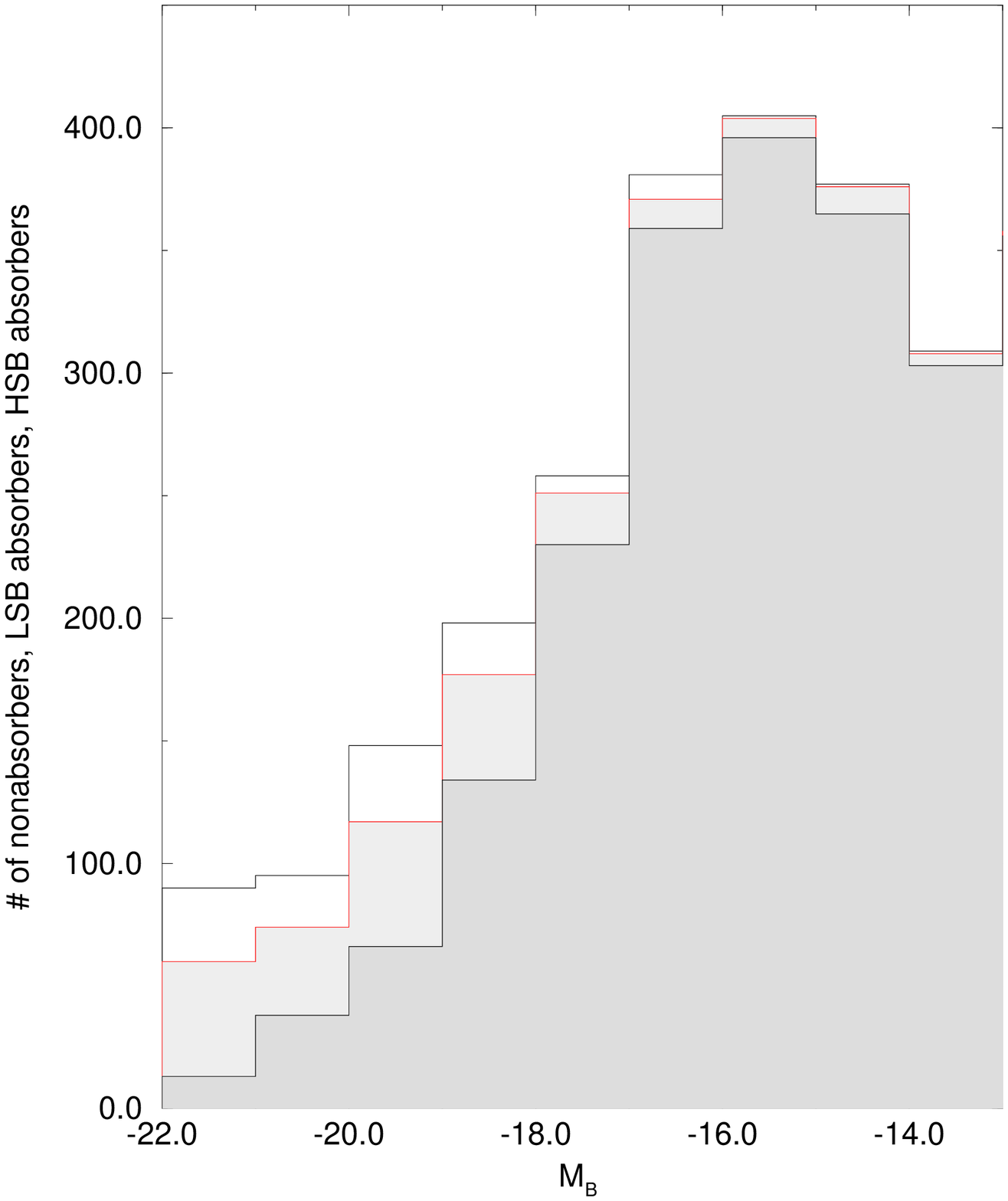}{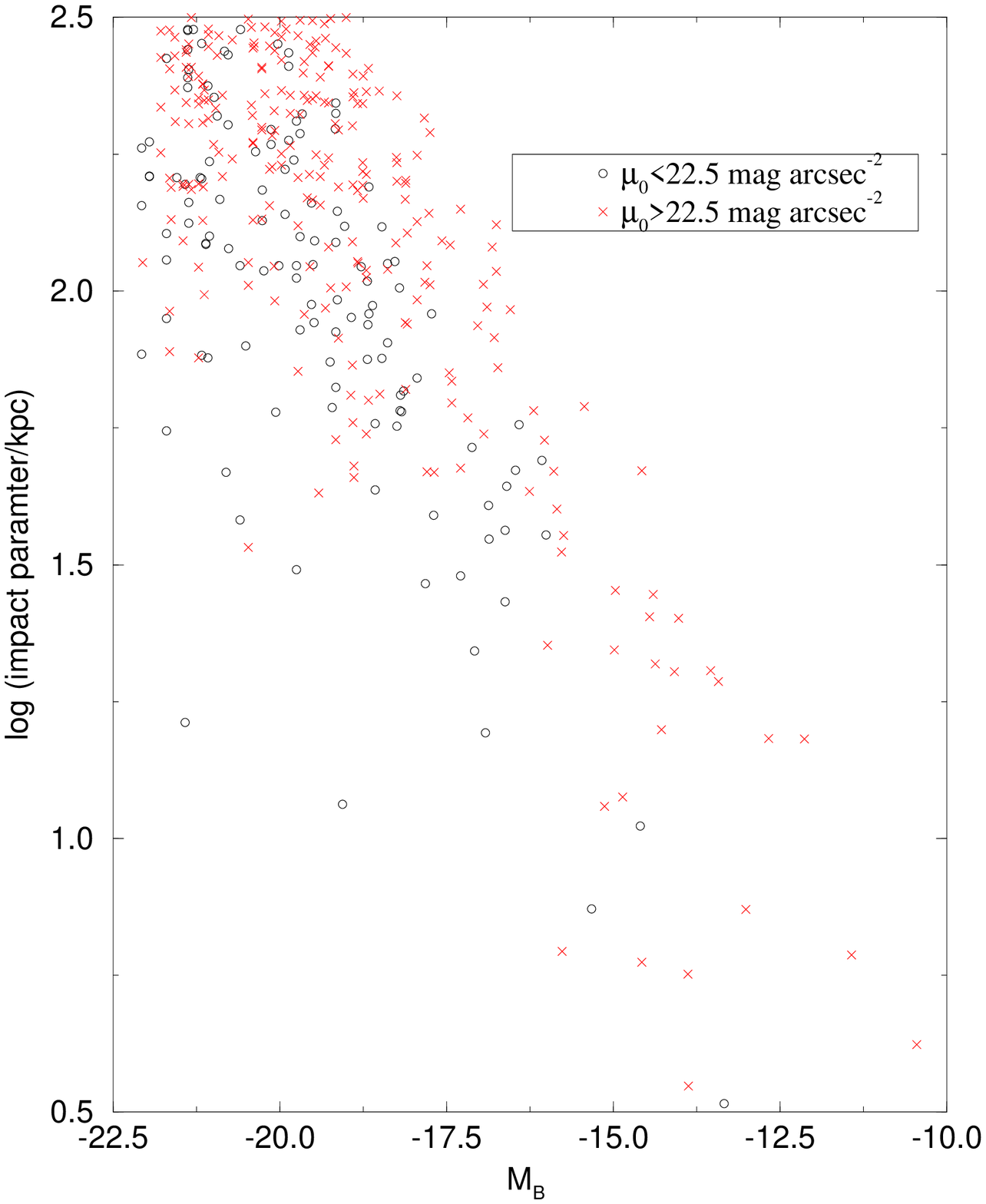}
\caption{a) The regions show (bottom to top) galaxies within 300 kpc of a line
of sight with no absorption $>10^{15}$ cm$^{-2}$,
absorption where $\mu_0>22.5$ B mag arcsec$^{-2}$,
and absorption for higher surface brightnesses.  
b) Impact parameter versus B magnitude for absorption $>10^{15}$ 
cm$^{-2}$.  Absorption at low impact parameters is dominated by low surface
brightness dwarfs which are unlikely to be identified.}\label{fig-2}
\end{figure}

Galaxies at impact parameters $<300$ kpc are binned by magnitude in Figure
~\ref{fig-2}a.  
Note that while dwarfs are numerous and 
usually cause no absorption,
a substantial fraction of absorbers at low impact parameters are dwarfs.  
Both dwarf and LSB absorbers are especially likely to be observationally
matched with the wrong galaxy, or no galaxy.  As in fig. 4a of Bowen et al.
(1996), the most absorption occurs at $M_B\sim -19$.  Dwarfs cause some
absorption, while luminous galaxies can be nonabsorbers due to
inclination.  Since luminous galaxies are the most extended
they may also be most affected by tidal interactions.
Figure~\ref{fig-2}b shows that dwarfs dominate absorption 
at low impact
parameters, simply due to the relative lack of luminous galaxies.
A dwarf absorber was discovered recently by van Gorkom et al. (1996).
Thus absorbers may be outer galaxy disks, but those closest to lines
of sight will be difficult to identify.

\end{document}